# Refraction and rightness in photonic crystals

R. Gajić[1,2], R. Meisels[2], F. Kuchar[2], K. Hingerl[3]

[1] *Institute of Physics, P.O.Box 68, 11080, Belgrade, Serbia*
[2] *Institute of Physics, University of Leoben, Franz Josef Str. 18, A-8700 Leoben, Austria*
[3] *Christian Doppler Lab, Institute for Semiconductors and Solid State Physics, University of Linz, Altenbergstrasse 69, A-4040 Linz, Austria*
*rgajic@phy.bg..ac.yu*

**Abstract:** We present a study on relation between the refraction and rightness effects in photonic crystals applied on a 2D square lattice photonic crystal. The plane wave (the band and equifrequency contour analyses) and FDTD calculations for both TM and TE modes revealed all possible refraction and rightness cases in photonic crystal structures in the first three bands. In particular, we show for the first time, a possibility of the left-handed positive refraction. This means that left-handedness does not necessarily imply negative refraction in photonic crystals.

## 1. Introduction

Recently left-handed materials attracted a lot of attention. These structures can be either realized as so-called meta-materials or photonic crystals. The left-handed materials were predicted in the seminal work by Veselago [1] who analyzed a hypothetical meta-material that had both the electrical permittivity ε and the magnetic permeability μ simultaneously negative. Veselago found out that the index of refraction is negative and that the Poynting vector **S** and the wave vector ***k*** (phase velocity) are anti-parallel. Since **E**, **H** and ***k*** form a left-handed set Veselago called the new materials 'left-handed' (LHM). In spite of some early efforts to fabricate LHMs, Veselago's work has stayed just an interesting idea for more than 30 years. In 1999, Pendry [3] proposed and tested the new meta-materials consisting of conducting loops (split-ring resonators (SRR)) or tubes ('Swiss rolls'). Their magnetic permeability μ has a resonance and a narrow frequency range with μ < 0. Shelby [4] reported the first LHM structure based on SRRs interconnected with a set of metallic rods (ε < 0). A prism made of this meta-material showed negative refraction at 10.5 GHz.

In contrast to the split rings or 'Swiss roll' structures, photonic crystals (PhC) consist of periodically modulated dielectric or metallic material. In the first case, the PhCs have modulated dielectric constant and always μ > 0. Nevertheless, it was shown that diffraction effects can produce different effects like 'super prism' [5], effective negative refraction or even negative index as in Veselago's metamaterials [6-8]. The losses can be much smaller in PhCs comparing to LHMs since a non-conducting dielectric material is used. Both negative refraction and left-handedness in PhCs were experimentally demonstrated in the microwave range [9-13]. The refraction experiments were performed on 2D dielectric [9-12] and metallic PhCs [13] by measuring the displacement of the transmitted beam at varying angle of incidence.

In this communication, we investigate the relation between the refraction sign and rightness in the case of 2D square lattice dielectric PhCs (see Figs 1 and 8). As tools, the plane wave method (PWM) and finite-difference time-domain (FDTD) calculations were used to analyze electromagnetic wave (EMW) propagation in the three lowest bands. The right-handed (RH) negative refraction has already been discussed [8] in the valence band (the first band below the gap) and here we show that the left-handed (LH) negative refraction is also possible in the second photonic band ("conduction" band) close to the second gap.

In the calculations, we use a 2D square lattice PhC made of $Al_2O_3$ rods in air. The ratio $r/a$ = 0.33 as we used earlier when we experimentally verified negative refraction at millimeter waves in the range 26 to 40 GHz [10, 12]. The propagation of EMW across both the ΓM and ΓX interface of the 2D square lattice PhC is investigated.

## 2. Basic notions

The structures showing negative refraction can be either the LH or RH media. (as in Fig. 1). The necessary condition for a LH PhC is $v_{ph} \cdot v_{gr} < 0$ or equivalently $k \cdot S < 0$, since $v_{gr}$ is parallel to **S** in PhCs large enough [7b]. The symbols $v_{ph}$, $v_{gr}$ and **S** indicate the phase and group velocity and the Poynting vector. The above condition will be used to distinguish the RH and LH behavior of the wave propagation in PhCs. In the case of the RH negative refraction (Fig. 1b) $v_{ph} \cdot v_{gr} > 0$ with $v_{gr}$ pointing to the 'negative' direction. Strictly speaking, Snell's law is defined only for the phase index of refraction $n_{ph}$ and the LH case exists if and only if $n_{ph}$ is negative. If the most of the propagating energy goes into the zero-order diffraction beam [8], one can apply Snell's law for an effective index of refraction: $n_{eff}(f, ki) \cdot \sin(\theta_i) = n_{air} \cdot \sin(\theta_r)$ where, $k_i$, $f$, $\theta_i$, and $\theta_r$ are the incident wave vector, the frequency, the incident and refracted angle, respectively. Since the PhC interfaces are chosen along symmetry directions, the equifrequency contour calculations (EFC) in the first Brillouin zone are sufficient to determine refraction properties of PhCs [7b]. Following [8-9] the necessary conditions for negative refraction require that EFCs in PhCs are both convex (inward gradient) and larger than the corresponding EFCs in air. Additionally $\lambda_o \equiv c/f \geq 2 \cdot a_s$ ($a_s$ is the surface lattice constant, $a_s = a \cdot \sqrt{2}$ for a square lattice PhC) in order to avoid higher order Bragg diffractions out of the crystal [7b]. In the first two bands of a PhC, with the parameters given above, this condition is fulfilled in the valence band (frequencies below 40 GHz: $\lambda_o \geq 2.85 \cdot a_s$) and is a fairly good approximation in the conduction band (below 68 GHz, $\lambda_o \geq 1.7 \cdot a_s$). The RH negative refraction can be observed around the M point in the Brillouin zone for both the TM and TE modes whereas the LH behavior is possible only in higher bands around the Γ point as it is shown in Fig. 2.

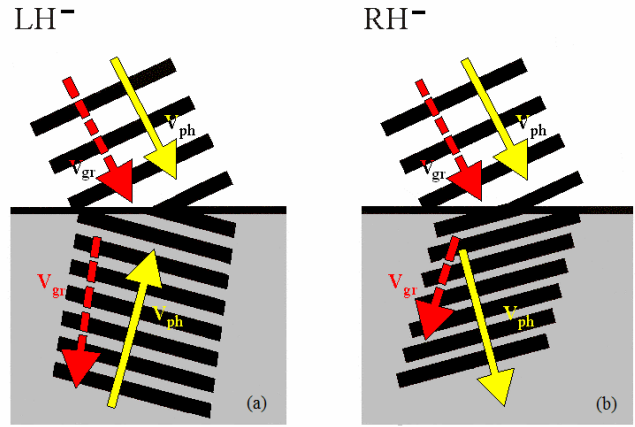

Fig. 1. LH⁻ (a) and RH⁻ (b) refraction (the sign ⁻ denotes negative refraction). For LH⁻, $v_{ph} \cdot v_{gr} < 0$ whereas for RH⁻, $v_{ph} \cdot v_{gr} > 0$.

## 3. Results

The band (Fig. 2) and EFC calculations, based on PWM using BandSOLVE [14], were performed for the first few TM/TE bands.

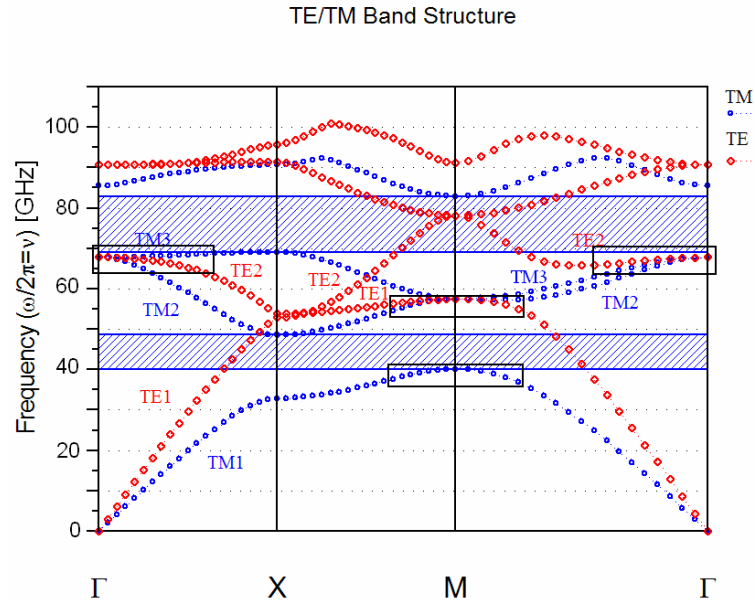

Fig. 2. Band structure of a 2D PhC made of $n = 3.1$ rods. The boxes determine the regions where negative refraction occurs.

The EFC plot for the first TM band the square lattice PhC is shown in Fig. 3. $k_\parallel$ is the parallel component of the $k$ vector, which is conserved in refraction. The EFCs are convex in the vicinity of the M point having an inward-pointing group velocity $v_{gr}$ [8]. Choosing a certain frequency of the incident EMW we construct the air EFC (the red circle). The red arrow indicates the incident TM EMW (**E** parallel to the rods) on the ΓM interface at an angle of 45° with the frequency of 36 GHz. Using the conservation of $k_\parallel$ we find the wave vector in the PhC, $k_{PhC}$. An inward gradient on the 36 GHz EFC in PhC (blue dashed arrow) determines the group velocity $v_{gr}$. From the directions of $k_{PhC}$ and $v_{gr}$, we conclude that the phase index $n_{ph} > 0$ and $n_{eff} < 0$ giving negative refraction without the LH effect in Fig.1. From Fig. 3 one can see that the air EFC is of the same size as the PhC one enabling an almost all-angle negative refraction. Due to similar effects around the M point in the Brillouin zone for the ΓM interface the TE1 EMW exhibits also the RH negative refraction for small incident angles.

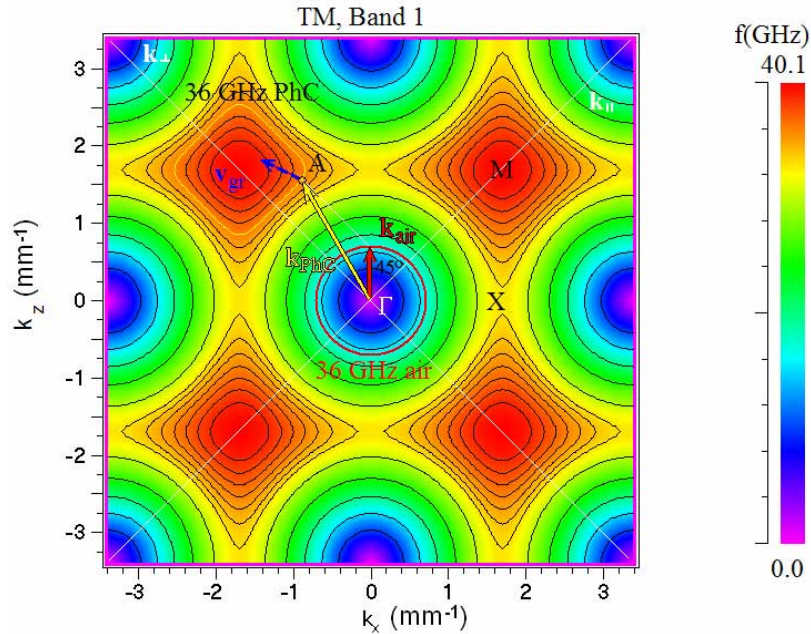

Fig. 3. EFC plot of the TM1 for a 2D PhC made of rods with $n = 3.1$. EFCs are presented for 10, 15, …,30 and 33, 34, …39 GHz.

For the determination of EMW propagation through the PhC, we performed FDTD simulations with FullWAVE [15]. Figure 4 shows the propagating pattern of the 36 GHz TM1 mode in the valence band.

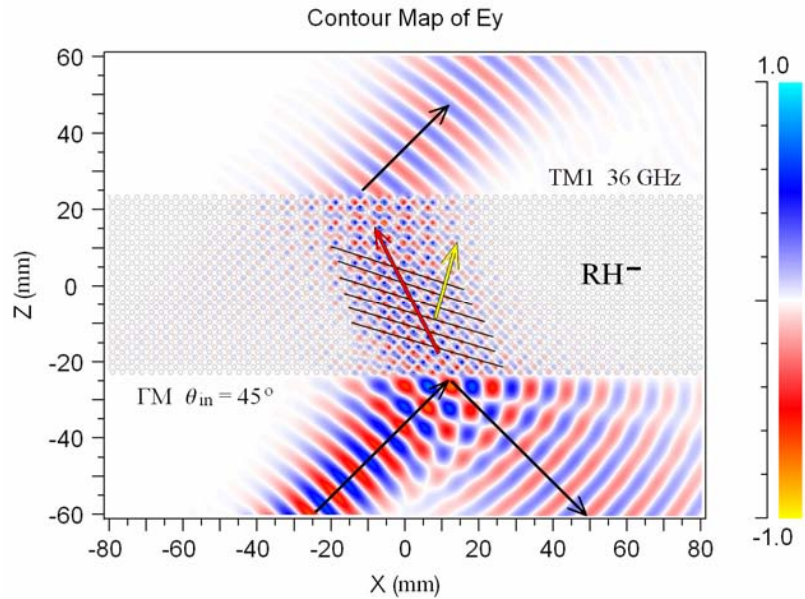

Fig. 4. Wave pattern of the TM1 wave incident at 45° on the ΓM interface of the PhC (FDTD) at $f = 36$ GHz. The black lines represent the wave front in the PhC. The red and yellow arrows denote $v_{gr}$ and $v_{ph}$, respectively.

The black lines represent the wave fronts in the PhC indicating that the phase velocity $v_{ph}$ and index $n_{ph}$ are > 0. The results of the EFC and FDTD calculations point clearly to the RH⁻ refraction as in Fig. 1b. In our previous papers [10, 12] we reported the transmission measurements at microwave frequencies between 26 and 40 GHz. For the TM1 EMW incident at 30 and 45° on the ΓM interface of the PhC negative refraction was observed in the range 35 to 38 GHz. The experimental findings were in a good agreement with the calculations.

In the calculations, the index of refraction of rods is taken to be 3.1. We propose this value as a correct one according to the transmission measurements we performed on the bulk alumina disc (diameter = 50 mm and thickness = 24 mm) in the Ka band (26.5-40 GHz). The average distance between interference fringes gave $n \approx 3.1$. This value is smaller than the one (3.3) used in [16] for $f > 20$ GHz and agrees with the value used in [9].

In the case of the TM2/TE2 bands around the Γ point, the EFCs are convex with an inward gradient. For the TE2 band the convex contours exist too and there is negative refraction for the incident angles less than 45° (ΓM interface) and 30° (ΓX interface) in the range 66-68 GHz. Here the LH⁻ refraction takes place for an incidence across both the ΓM and ΓX interface of the 2D square lattice PhC as it is shown in Fig. 5. There, the negative refraction of a 67 GHz TE polarized beam at an incident angle of 15° for both the ΓM and ΓX surface is analyzed. We find degeneracy for an incidence on the ΓM interface. For small incident angles, both the RH⁻ and LH⁻ waves are simultaneously excited as shown in Fig. 5 by the vectors, ending in point B for the LH case and in C for the RH one. They are distinguished by the sign of $v_{ph} \cdot v_{gr}$. Unfortunately, the RH⁻ and LH⁻ beams have almost the same direction and are not resolved in the FDTD simulations. For an incidence across the ΓX interface, there is just the LH⁻ wave in the PhC.

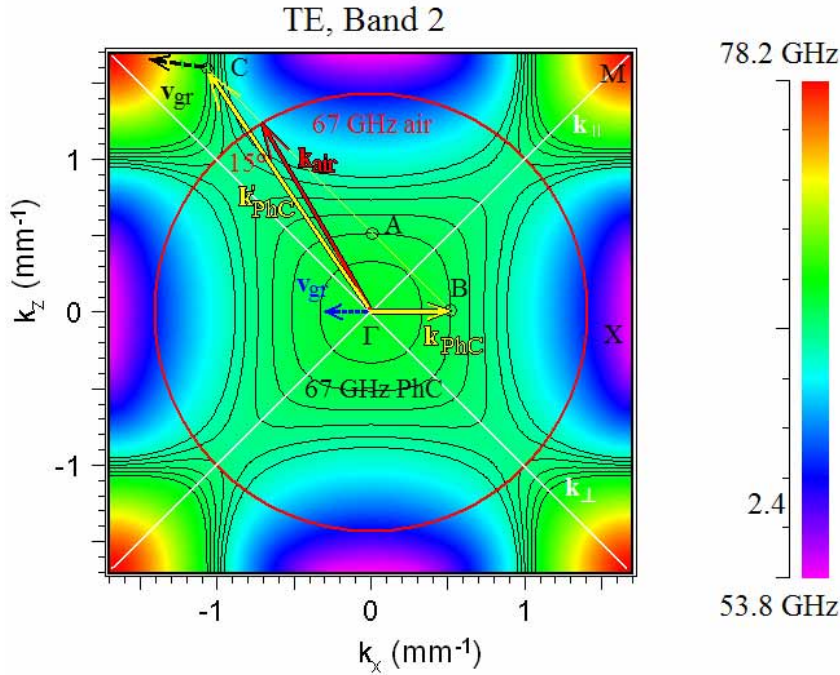

Fig. 5. EFC plot for the second TE band of a 2D square lattice PhC of $n$ =3.1 rods in the case of an EMW incidence of 15° in respect to the normal to the ΓM interface. The red and yellow arrows denote the incident EMW of 67 GHz and the wave vectors ($k_{PhC}$ and $k'_{PhC}$), respectively. The dashed arrows represent group velocities. EFCs are shown with 0.5 GHz steps between 65 and 67.5 GHz. The wave with $k_{PhC}$ at point A does not exist since the corresponding $v_{gr}$ would not give an energy flow away from the source.

In both cases (Fig. 5), there is the LH⁻ refraction with vph and $v_{gr}$ being nearly anti-collinear like in the Veselago's metamaterials.

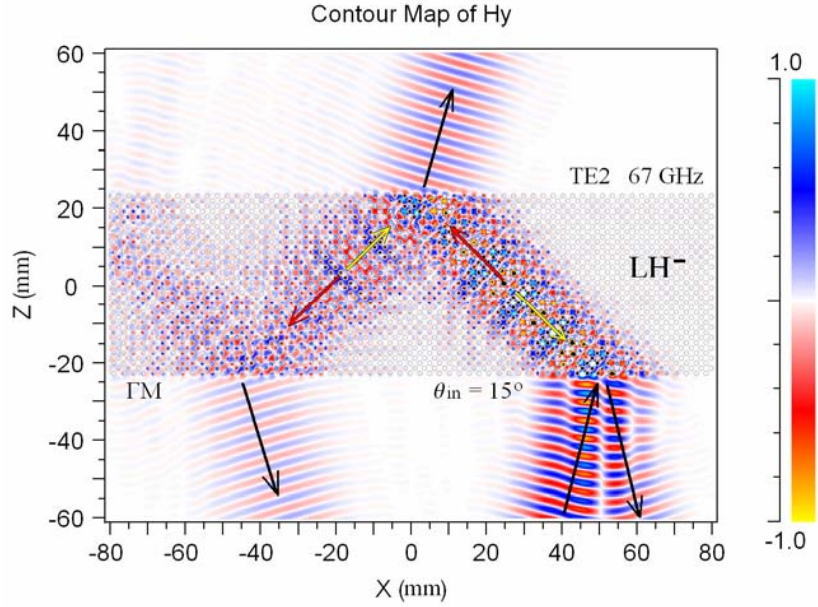

Fig. 6. Propagation wave pattern of the 67 GHz TE2 wave at an incidence of 15° across the ΓM interface of the PhC (FDTD simulation) exhibiting the Veselago LH⁻ refraction. The red and yellow arrows denote $v_{gr}$ and $v_{ph}$, respectively.

Figures 6 and 7 show propagation wave patterns for an incidence at the ΓM and ΓX interface, respectively. In agreement with the EFC calculation in Fig. 5, the EMW front patterns point to left-handedness as depicted in Fig. 1a. In contrast to negative refraction in the valence band, the propagating states here are close to Γ and have small group velocities.

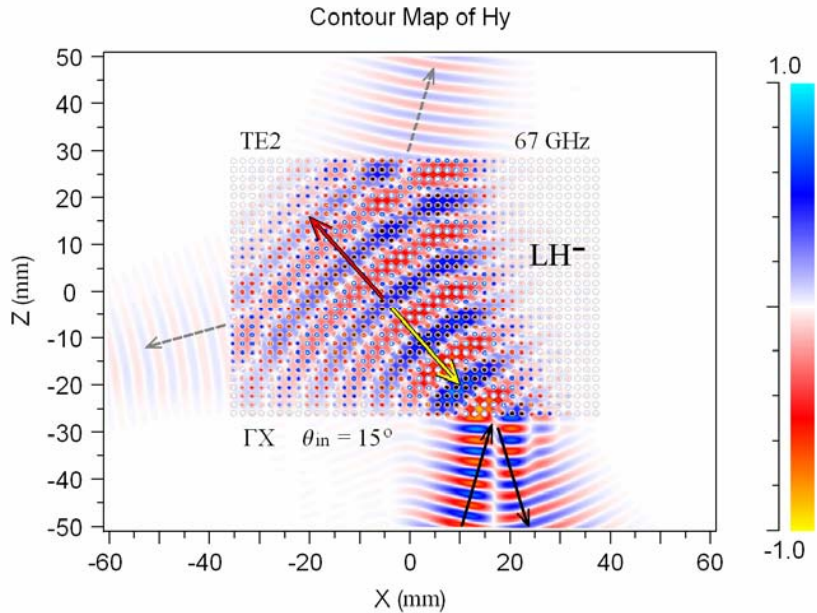

Fig. 7. Propagation wave pattern of the 67 GHz TE wave at an incidence of 15° on the ΓX interface of the PhC (FDTD) exhibiting the Veselago LH⁻ refraction. The dashed gray outgoing arrows denote the temporary positions of the beams. The red and yellow arrows denote $v_{gr}$ and $v_{ph}$, respectively, as before.

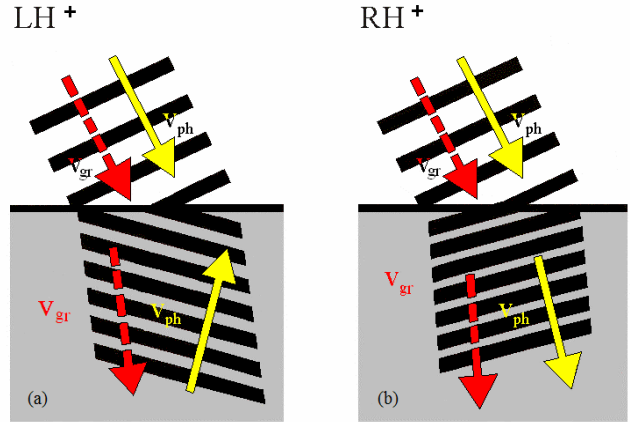

Fig. 8. Similarly as in Fig. 1, positive refraction could be realized as a LH+ (a) or a RH+ (b) beam.

In addition, the LH beam in the TE2 band occurs near the Γ point close to the partial TE gap in the ΓX direction (see Fig. 2). Our PhC has no full TE band-gaps, but a partial gap along the ΓX direction, which affects the EFCs around the Γ point making them round with inward gradients. The TM2 band has also convex contours with negative gradient around Γ between around 65 and 68 GHz.

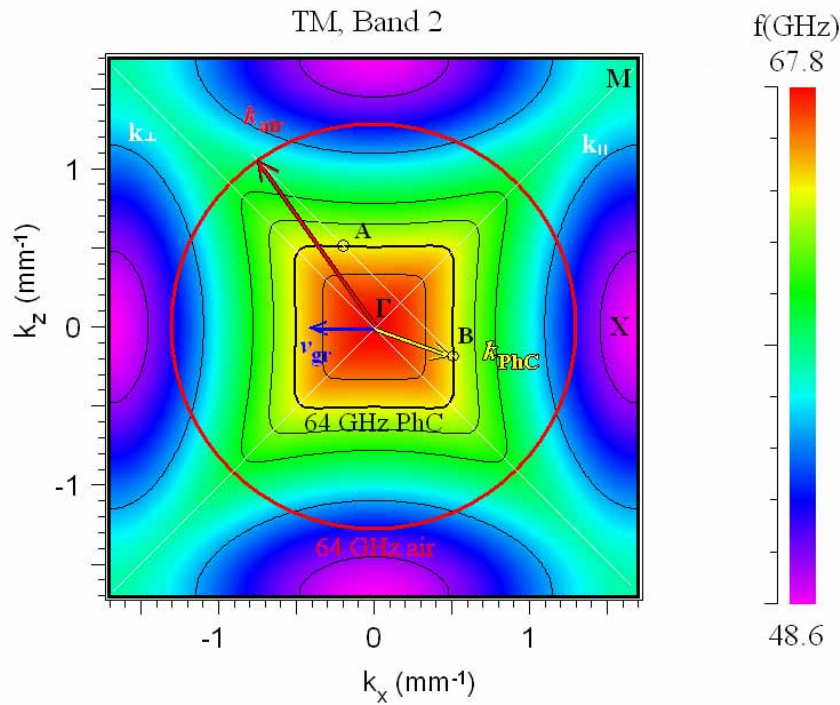

Fig. 9. EFC plot of the TM2 band of a 2D square lattice PhC of $n = 3.1$ rods. The red circle denotes the 64 GHz EFC in air. The PhC EFCs are shown at 50, 55, 60, 62, 64 and 66 GHz. The point B corresponds to an outgoing wave as before.

Comparing to the TE2 band the TM2 EFCs are square-like with rounded corners for most frequencies as in Fig. 9. For an angle of incidence of 10° at the ΓM interface, the collimated LH⁻ beam is obtained as in Fig. 11. The same figure reveals a second beam excited in the TM3 band directed in positive direction. For the frequencies up to 67 GHz (Fig. 11), there is a LH+ refraction! The inner product $v_{ph} \cdot v_{gr} < 0$ as in Fig. 10 and depicted in Fig. 8a. Fotinopoulou [7] showed that in the case of normal dispersion (when $n_{ph}$ is independent of the incident angle) $n_{ph}$ and $n_{gr}$ have the same sign and LH appears if and only if $n_{ph} < 0$. For anisotropic dispersion, as in our case, it is not valid any more. This means that left-handedness is not necessarily related to negative refraction.

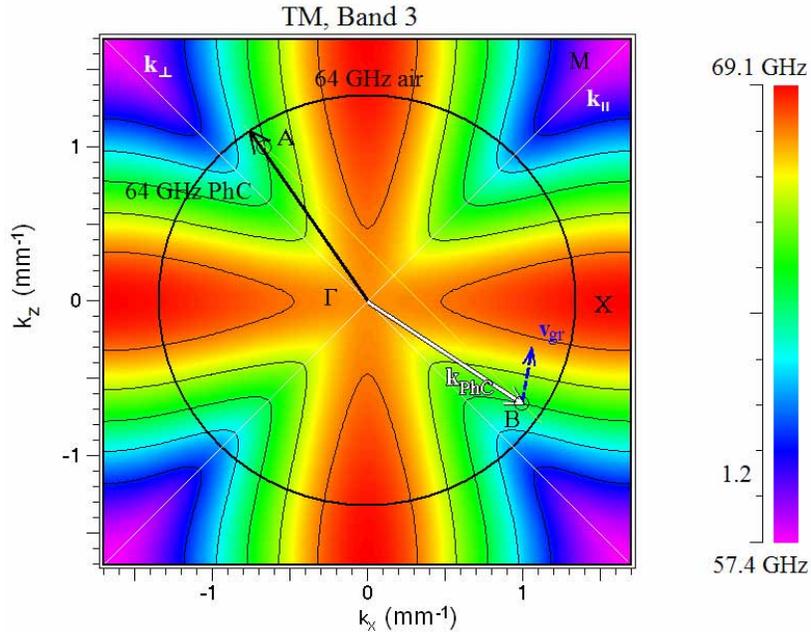

Fig. 10. EFC plot for the TM3 band of a 2D square lattice PhC made of $n = 3.1$ rods in the case of the 64 GHz EMW incident at 10° across the ΓM interface. The black and white arrows denote the incident air wave vector and phase velocity in the crystal ($k_{PhC}$), respectively. The dash blue arrow represents the group velocity. The EFCs are shown in the range 60 to 68 GHz with 2 GHz steps.

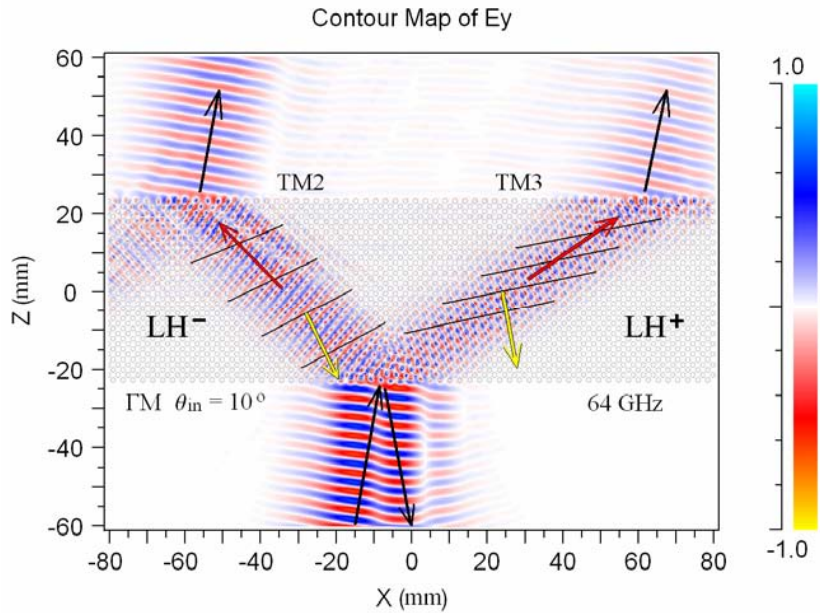

Fig. 11. Propagation wave pattern of the 64 GHz TM wave at an incidence of 10° on the ΓM interface of the PhC. Both beams are LH whereas just the left one (from the TM2 band) exhibits the LH negative refraction. The right beam corresponds to the TM3 band and shows coexistence between LH and positive refraction as in Fig. 8. $LH^{\pm}$ refers to positive (negative) LH refraction. The black lines represent the wave fronts of the beams. The red and yellow arrows denote $v_{gr}$ and $v_{ph}$ as defined in Figs. 1 and 8.

The directions and magnitudes of $k_{ph}$ and $k_{gr}$ in the PHC were inferred from both the EFC and FDTD calculations. There is an excellent agreement between two methods. For instance, using Figs 9-11, the obtained angles of $k_{ph}^{(EFC)}$ and $k_{ph}^{(FDTD)}$ for the LH⁻ and LH⁺ beams do not differ by more than 1°, whereas the difference of their magnitudes is less than 1 %. In all cases with the LH$^\pm$ behavior we noticed the backward waves in the FDTD simulations (see the movies in Figs. 12 and 13).

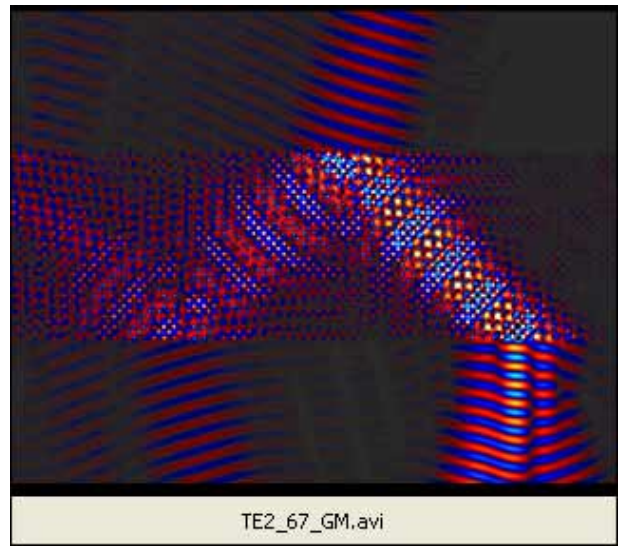

Fig. 12. Backward waves of the 67 GHz TE2 mode (FDTD) as in Fig. 6 exhibiting the Veselago LH$^-$ refraction

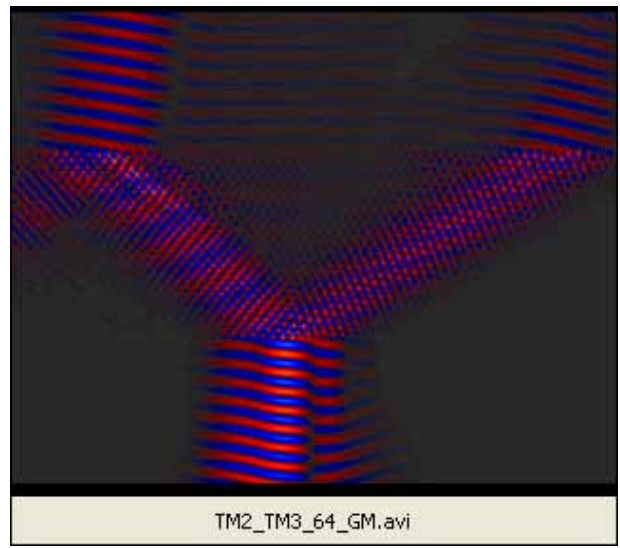

Fig. 13. Backward waves in the LH$^-$ (left, TM2) and LH$^+$ beams (right, TM3) as in Fig. 11.

## 4. Summary

We analyzed all refraction and rightness combinations in 2D square lattice PhC. In the valence band for both TM1 and TE1 modes there is a RH$^-$ refraction due to the inward EFC gradients close to the M point. On the other hand, for the TM2 and TE2 bands around the Γ point, we found the Veselago LH$^-$ cases where $v_{ph}$ and $v_{gr}$ are almost anti-collinear due to the presence of either the full or the partial gap. Especially interesting is the case of the LH$^+$ refraction in the TM3 band, which has not been reported earlier proving that left-handedness is not necessary connected to negative refraction in PhCs

**Acknowledgments**

R.G. acknowledges the support of the Serbian Ministry of Science and Environment Protection. The Christian Doppler Laboratory is grateful to Photeon and Dr. Heinz Syringe for financial support and to Dr. Johann Messer from the Linz Supercomputer Center.